\documentclass[apjl]{emulateapj}

\def\kms{\ifmmode{\rm km\thinspace s^{-1}}\else km\thinspace s$^{-1}$\fi}
\def\ms{\ifmmode{\rm m\thinspace s^{-1}}\else m\thinspace s$^{-1}$\fi}
\def\gschat3{GSC~03466-00819}
\def\hatp3{HAT-P-3}
\newcommand{\rhk}{\ensuremath{R^{\prime}_{\rm HK}}}
\newcommand{\logrhk}{\ensuremath{\log\rhk}}

\shortauthors{Torres}
\shorttitle{HAT-P-3b}

\begin{document}

\title{HAT-P-3\lowercase{b}: A heavy-element rich planet transiting a
K dwarf star}

\author{
G.\ Torres\altaffilmark{1},
G.\ \'A.\ Bakos\altaffilmark{1,2},
G.\ Kov\'acs\altaffilmark{3},
D.\ W.\ Latham\altaffilmark{1},
J.\ M.\ Fern\'andez\altaffilmark{1},
R.\ W.\ Noyes\altaffilmark{1},
G.\ A.\ Esquerdo\altaffilmark{1},
A.\ Sozzetti\altaffilmark{1,4},
D.\ A.\ Fischer\altaffilmark{5},
R.\ P.\ Butler\altaffilmark{6},
G.\ W.\ Marcy\altaffilmark{7},
R.\ P.\ Stefanik\altaffilmark{1},
D.\ D.\ Sasselov\altaffilmark{1},
J.\ L\'az\'ar\altaffilmark{8},
I.\ Papp\altaffilmark{8},
P.\ S\'ari\altaffilmark{8}
}

\altaffiltext{1}{Harvard-Smithsonian Center for Astrophysics,
Cambridge, MA, gtorres@cfa.harvard.edu}

\altaffiltext{2}{Hubble Fellow}

\altaffiltext{3}{Konkoly Observatory, Budapest, P.O.\ Box 67, H-1125,
Hungary}

\altaffiltext{4}{INAF - Osservatorio Astronomico di Torino, 10025 Pino
Torinese, Italy}

\altaffiltext{5}{Department of Physics and Astronomy, San Francisco
State University, San Francisco, CA}

\altaffiltext{6}{Department of Terrestrial Magnetism, Carnegie
Institute of Washington, DC}

\altaffiltext{7}{Department of Astronomy, University of California,
Berkeley, CA}

\altaffiltext{8}{Hungarian Astronomical Association, Budapest,
Hungary}

\begin{abstract} 

We report the discovery of a Jupiter-size planet transiting a
relatively bright ($V = 11.56$) and metal-rich early K dwarf star with
a period of $\sim2.9$ days. On the basis of follow-up photometry and
spectroscopy we determine the mass and radius of the planet, \hatp3b,
to be $M_p = 0.599 \pm 0.026$~M$_{\rm Jup}$ and $R_p = 0.890 \pm
0.046$~R$_{\rm Jup}$. The relatively small size of the object for its
mass implies the presence of about 75~M$_{\earth}$ worth of heavy
elements ($\sim\frac{1}{3}$ of the total mass) based on current
theories of irradiated extrasolar giant planets, similar to the mass
of the core inferred for the transiting planet HD~149026b. The bulk
density of \hatp3b\ is found to be $\rho_p = 1.06 \pm 0.17$
g~cm$^{-3}$, and the planet orbits the star at a distance of 0.03894
AU. Ephemerides for the transit centers are $T_c = 2,\!454,\!218.7594
\pm 0.0029 + N\times (2.899703 \pm 0.000054)$ (HJD).
		
\end{abstract}


\section{Introduction}
\label{sec:introduction}

Ground-based surveys for extrasolar transiting planets have produced
18 discoveries to date. This sample is still small enough that
individual discoveries often advance our understanding of these
objects significantly by pushing the limits of parameter space, either
in planet mass, radius, or some other property. Here we announce a new
transiting planet, the third to come out of the HATNet survey, around
a star previously known as \gschat3. It is the smallest yet discovered
photometrically, and appears to have a heavy-element content
representing a substantial fraction of its mass.
	
\section{Photometric detection}
\label{sec:detection}

% Alternate names
%
% TYC 3466-819-1
% 2MASS 13442258+4801432
% GSC 03466-00819
% SDSS J134422.59+480143.1

\begin{figure} 
%\vskip -1in
\epsscale{1.0} 
\plotone{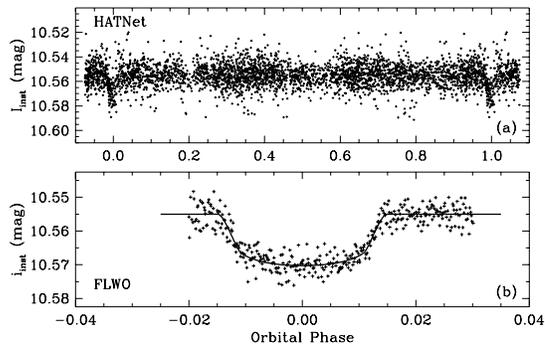}
\vskip -0.8in 
\figcaption[]{(a) Unbinned instrumental $I$-band
discovery light curve of \hatp3\ obtained with HATNet, folded with the
period of $P = 2.899703$ days. (b) Unbinned instrumental Sloan
$i$-band photometry collected on UT 2007 April 27 with the 1.2-m
telescope at FLWO, along with our best-fit transit model curve (see
text).\label{fig:lc}}
\end{figure}

\gschat3\ is located in our internally labeled field ``G145'' centered
at $\alpha = 13^{\rm h} 48^{\rm m}$, $\delta = +45\arcdeg 00\arcmin$,
which was observed using the 11-cm aperture HAT-5 instrument of the
HAT network \citep[HATNet;][]{Bakos:02, Bakos:04}, located at the F.\
L.\ Whipple Observatory (FLWO) on Mt.\ Hopkins (Arizona). The field
was observed in single-station mode, i.e., without networked
operations between Arizona and HATNet's other station in
Hawaii. Observations over an extended interval of six months (from
2006 January 28 through 2006 July 22) yielded some 3200 photometric
measurements per star for approximately $10,\!000$ stars in the field
down to $I\sim14$.  Following standard frame calibration procedures
astrometry was performed as described in \citet{Pal:06}, and aperture
photometry results were subjected to the de-trending algorithms we
refer to as External Parameter Decorrelation \citep[EPD, described
briefly in][]{Bakos:07}, and Trend Filtering Algorithm
\citep[TFA;][]{Kovacs:05}. We searched the light curves for box-shaped
transit signals using the BLS algorithm of \citet{Kovacs:02}. A
prominent 2.9-day signal was detected in the $I\approx 10.52$
magnitude star \gschat3 (also known as 2MASS 13442258+4801432; $\alpha
= 13^{\rm h} 44^{\rm m} 22\fs58$, $\delta = +48\arcdeg 01\arcmin
43\farcs2$; J2000) with a transit depth of $0.014$ mag and a
dip-significance parameter (DSP) of $\sim11$ \citep{Kovacs:05b}. This
discovery light curve is shown in Figure~\ref{fig:lc}a. The standard
deviation of the measurements after the EPD procedure was 0.009
mag. Altogether 5 partial and 2 full transits were observed.  As is
shown in the following sections the transiting companion is a Jovian
planet.  Hereafter we refer to the star as HAT-P-3, and to the
planetary companion as HAT-P-3b.
The near infrared color of the star ($J\!-\!K_s = 0.49$) corresponds
to an early K dwarf, and the annual proper motion listed in the
Tycho-2 catalog \citep[$\mu_{\alpha}\cos\delta = -23.3 \pm 2.9$
mas~yr$^{-1}$, $\mu_{\delta} = -24.0 \pm 2.6$
mas~yr$^{-1}$;][]{Hog:00} is consistent with this.  The photometric
ephemeris was determined by combining all recorded HATNet transit
events with a full transit observed more recently at higher precision
with a larger telescope, as described below. The period we obtain is
$P = 2.899703 \pm 0.000054$ days, and the reference epoch of transit
center is $T_c = 2,\!454,\!218.7594 \pm 0.0029$ (HJD). We use this
ephemeris for the remainder of the paper.

We note that HAT-P-3 has a faint ($I\approx15$) and red ($J\!-\!K_s =
0.86$) companion at an angular separation of $\sim$10\arcsec\ (2MASS
13442345+4801387), which is entirely within the aperture used for the
HATNet photometry. If this happened to be an eclipsing binary, we
estimate, based on the relative fluxes, that 1.6-mag deep eclipses of
this star could mimic the observed shallow transits seen in the merged
light curve. However, our follow-up photometry described in
\S\ref{sec:followup} with much better spatial resolution shows that
the shallow dips in brightness are indeed present in the light curve
of HAT-P-3, whereas the faint companion is constant at the 0.05 mag
level.

\section{Follow-up observations}
\label{sec:followup}

\hatp3\ was observed spectroscopically with the FLWO 1.5-m Tillinghast
reflector in order to rule out the possibility that the observed drop
in brightness is caused by a stellar companion in an eclipsing
binary. Three spectra were obtained with the CfA Digital Speedometer
\citep{Latham:92} over an interval of 25 days. These observations
cover 45~\AA\ in a single echelle order centered at 5187~\AA, and have
a resolving power of $\lambda/\Delta\lambda \approx 35,\!000$.  Radial
velocities were obtained by cross-correlation and have a typical
precision of 0.4~\kms. They showed no variation within the
uncertainties, ruling out a companion of stellar mass. The mean
velocity is $-23.8 \pm 0.1$~\kms.

Higher resolution spectroscopy was conducted with the HIRES instrument
\citep{Vogt:94} on the Keck~I telescope. With a spectrometer slit of
$0\farcs86$ the resolving power is $\lambda/\Delta\lambda \approx
55,\!000$, and the wavelength coverage is $\sim$3800--8000~\AA. An
iodine gas absorption cell was used to superimpose a dense forest of
$I_2$ lines on the stellar spectrum and establish a highly accurate
wavelength fiducial \citep[see][]{Marcy:92}. A total of nine 12-minute
exposures were obtained between 2007 March and June with the iodine
cell, along with one without $I_2$ for use as a template. Relative
radial velocities in the Solar System barycentric frame were derived
as described by \cite{Butler:96}, including full modeling of the
spatial and temporal variations of the instrumental profile. The
results, listed in Table~\ref{tab:rvs}, have internal errors
$\lesssim 3~\ms$.
\begin{deluxetable}{lccc}
\tablewidth{0pc}
\tablecaption{Relative radial velocity measurements of \hatp3.\label{tab:rvs}}
\tablehead{\colhead{BJD} & \colhead{RV} & \colhead{$\sigma_{\rm RV}$} & \colhead{} \\
\colhead{\hbox{~~~~(2,400,000$+$)~~~~}} & \colhead{(\ms)} & \colhead{(\ms)} & \colhead{Phase}}
\startdata
 54187.01101\dotfill &  $-$38.9   &  1.5  & 0.051 \\
 54187.12931\dotfill &  $-$67.4   &  1.6  & 0.092 \\
 54187.97909\dotfill &  $-$78.9   &  1.8  & 0.385 \\
 54188.05795\dotfill &  $-$59.3   &  1.3  & 0.412 \\
 54188.98084\dotfill &  $+$75.9   &  1.4  & 0.730 \\
 54189.06253\dotfill &  $+$75.9   &  1.4  & 0.759 \\
 54189.11203\dotfill &  $+$73.9   &  1.5  & 0.776 \\
 54250.85697\dotfill &  $-$48.1   &  2.1  & 0.069 \\
 54278.86111\dotfill &  $+$65.2   &  2.9  & 0.727 \\
\noalign{\vskip -6pt}
\enddata
\end{deluxetable}

The iodine-free template spectrum from Keck was used to infer the
atmospheric parameters of the star. A spectral synthesis modeling was
carried out using the SME software \citep{Valenti:96}, with wavelength
ranges and atomic line data as described by \cite{Valenti:05}. The
effective temperature ($T_{\rm eff}$), surface gravity ($\log
g_\star$), iron abundance ([Fe/H]), and projected rotational velocity
($v \sin i$) we obtain are listed in Table~\ref{tab:stellar}, where
the uncertainties represent internal errors. The temperature and
surface gravity correspond to an early K dwarf.
\vskip -10pt
\begin{deluxetable}{lcl}
\tablewidth{0pc}
\tablecaption{Stellar parameters for \hatp3.\label{tab:stellar}}
\tablehead{\colhead{~~~~~~Parameter~~~~~~} & \colhead{Value} & \colhead{Source}}
\startdata
$T_{\rm eff}$ (K)\dotfill   &  5185~$\pm$~46\phn\phn & SME\tablenotemark{a} \\
$[$Fe/H$]$\dotfill          &  +0.27~$\pm$~0.04\phs  & SME \\
$\log g_\star$ (cgs)\dotfill    &  4.61~$\pm$~0.05       & SME \\
$v \sin i$ (\kms)\dotfill   &  0.5~$\pm$~0.5         & SME \\
$M_\star$ (M$_{\sun}$)\dotfill  &  $0.936_{-0.062}^{+0.036}$  & Y$^2$+LC+SME\tablenotemark{b} \\
$R_\star$ (R$_{\sun}$)\dotfill  &  $0.824_{-0.035}^{+0.043}$  & Y$^2$+LC+SME \\
$L_\star$ (L$_{\sun}$)\dotfill  &  $0.442_{-0.057}^{+0.078}$   & Y$^2$+LC+SME \\
$M_V$ (mag)\dotfill         &  5.86~$\pm$~0.20       & Y$^2$+LC+SME \\
Age (Gyr)\dotfill           &  $0.4_{-0.3}^{+6.5}$   & Y$^2$+LC+SME \\
Distance (pc)\dotfill       &  140~$\pm$~13\phn      & Y$^2$+LC+SME \\
\noalign{\vskip -5pt}
\enddata
\tablenotetext{a}{SME = `Spectroscopy Made Easy' package for analysis
of high-resolution spectra \cite{Valenti:96}. See text.}
\tablenotetext{b}{Y$^2$+LC+SME = Yale-Yonsei isochrones \citep{Yi:01},
light curve parameters, and SME results.\vskip 3pt}
\end{deluxetable}
\vskip 0.1in
Photometric follow-up of \hatp3\ was carried out in the Sloan $i$ band
with KeplerCam \citep[see][]{Holman:07} on the 1.2-m telescope at
FLWO. After several attempts thwarted by weather we obtained a
reasonably complete and high-precision light curve of \hatp3\ on UT
2007 April 27. An astrometric solution between the individual frames
and the 2MASS catalog was carried out using first order polynomials
based on $\sim$100 stars per frame. Aperture photometry was performed
using a series of apertures on fixed positions around the 2MASS-based
[$X$,$Y$] pixel coordinates. We selected a frame taken near the
meridian and used $\sim$70 stars and their magnitudes as measured on
this reference frame to transform all other frames to a common
instrumental magnitude system (weighting by the estimated photometric
errors of each star). The aperture yielding the lowest scatter outside
of transit was used in the subsequent analysis. The light curve of
HAT-P-3 (Figure~\ref{fig:lc}b) was then de-correlated against trends
using the out-of-transit sections and a dependence on hour angle and
zenith distance. The photometry is somewhat affected by occasional
thin cirrus during the night, and has a typical precision of
$\sim$2.5~mmag per individual measurement at 20-second cadence.

\section{Analysis}
\label{sec:analysis}

Modeling of the $i$-band light curve was carried out using the
formalism of \cite{Mandel:02}. Quadratic limb-darkening coefficients
($u_1 = 0.3832$ and $u_2 = 0.2683$) were taken from the tables of
\cite{Claret:04} by interpolation to the stellar properties given in
Table~\ref{tab:stellar}. The period was held fixed, and the orbit was
assumed to be circular, since the eccentricity is not sufficiently
well constrained by the presently available velocity measurements.
The four adjusted parameters are the planet-to-star ratio of the radii
($R_p/R_\star$), the normalized separation ($a/R_\star$) where $a$ is
the semimajor axis of the relative orbit, the impact parameter ($b
\equiv a \cos i/R_\star$), and the time of the center of the transit
($T_c$). A grid search was performed at a fixed $T_c$ to locate the
$\chi^2$ minimum in the parameter space of the other three variables,
and these were subsequently held fixed to optimize $T_c$. The process
was iterated until convergence. Uncertainties were estimated by
alternately fixing each parameter at a range of values near the
minimum and allowing the others to float, until the best-fit solution
produced an increase in the $\chi^2$ corresponding to a 1$\sigma$
change. The results are shown in Table~\ref{tab:parameters}.

\begin{deluxetable}{lc}
\tablewidth{0pc}
\tablecaption{Spectroscopic and light curve solutions for \hatp3, and
inferred planet parameters.\label{tab:parameters}}
\tablehead{\colhead{~~~~~~~~~~~~~~Parameter~~~~~~~~~~~~~~} & \colhead{Value}}
\startdata
\noalign{\vskip -4pt}
\sidehead{Spectroscopic parameters}
~~~$P$ (days)\tablenotemark{a}\dotfill              &  2.899703~$\pm$~0.000054 \\
~~~$T_c$ (HJD$-2,\!400,\!000$)\tablenotemark{a}\dotfill             & $54,\!218.7594$~$\pm$~0.0029\phm{,2222} \\
~~~$K$ (\ms)\dotfill               & 89.1~$\pm$~2.0\phn \\
~~~$\gamma$ (\ms)\dotfill          & $-$14.8~$\pm$~1.5\phn\phs \\
~~~$e$\dotfill                     & 0 (adopted) \\
\sidehead{Light curve parameters}
~~~$a/R_\star$\dotfill                 & $10.59_{-0.84}^{+0.66}$ \\
~~~$R_p/R_\star$\dotfill               & $0.1109_{-0.0022}^{+0.0025}$ \\
~~~$b \equiv a \cos i/R_\star$\dotfill & $0.51_{-0.13}^{+0.11}$ \\
~~~$i$ (deg)\dotfill               & 87.24~$\pm$~0.69\phn \\
~~~Transit duration (days)\dotfill & 0.0858~$\pm$~0.0020 \\
%~~~$u_1$\tablenotemark{b}\dotfill  & 0.3832 (adopted) \\
%~~~$u_2$\tablenotemark{b}\dotfill  & 0.2683 (adopted) \\
\sidehead{Planet parameters}
~~~$M_p$ (M$_{\rm Jup}$)\dotfill & 0.599~$\pm$~0.026 \\
~~~$R_p$ (R$_{\rm Jup}$)\dotfill & 0.890~$\pm$~0.046 \\
~~~$\rho_p$ (g~cm$^{-3}$)\dotfill  & 1.06~$\pm$~0.17 \\
~~~$a$ (AU)\dotfill              & 0.03894~$\pm$~0.00070 \\
~~~$\log g_p$ (cgs)\dotfill        & 3.310~$\pm$~0.066 \\
\noalign{\vskip -3pt}
\enddata
\tablenotetext{a}{Held fixed from the photometric determination (\S\ref{sec:detection}).}
\end{deluxetable}

\begin{figure} 
%\vskip -1in
\epsscale{1.05} 
\plotone{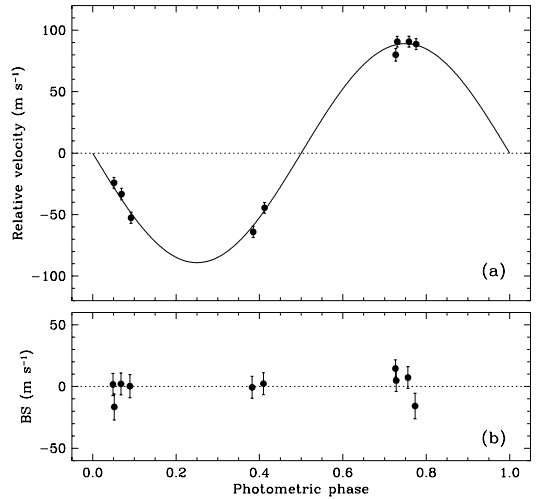}
%\vskip -0.1in 
\figcaption[]{(a) Radial-velocity measurements for
\hatp3\ along with our orbital fit, shown as a function of orbital
phase. The center-of-mass velocity has been subtracted. (b) Bisector
spans (BS) for our 10 Keck spectra (template + 9 iodine exposures),
computed as described in the text. The mean value has been subtracted,
and the vertical scale is the same as in the top
panel.\vskip 5pt\label{fig:rvbis}}
\end{figure}

The relative radial velocities of \hatp3\ were fitted with a Keplerian
model constrained to have zero eccentricity. The ephemeris was held
fixed at the value for $P$ determined in \S\ref{sec:detection} and
$T_c$ from the light curve analysis. The solution fits the data well
(see Figure~\ref{fig:rvbis}a), but has an rms scatter of 4.7~\ms\
(accounting for the number of degrees of freedom) that is larger than
the typical internal errors (1.3--2.9~\ms). No obvious long-term
trends are seen in the residuals. Inspection of the spectra reveals
that the \ion{Ca}{2} H and K lines display modest emission cores,
indicative of a small degree of chromospheric activity. This is known
to cause excess scatter in the velocities \citep[e.g.,][]{Saar:97},
and is often referred to as ``jitter''.  While we cannot rule out
other systematics that may be causing our internal errors to be
underestimated, we attribute most of the excess scatter to
activity. For our final fit we have added 4.2~\ms\ in quadrature to
our formal errors, which yields a reduced $\chi^2$ near unity. The
parameters of the fit are not significantly changed, and are listed in
Table~\ref{tab:parameters}.  The minimum companion mass is $M_p \sin i
= 0.625 \pm 0.015 \times [(M_\star + M_p)/M_{\sun}]^{2/3}$~M$_{\rm Jup}$,
where $M_\star$ represents the mass of the star.

We investigated the possibility that the radial velocities we measured
are the result of distortions in the line profiles due to
contamination from an unresolved eclipsing binary \citep{Santos:02,
Torres:05}, instead of being due to true Doppler motion in response to
a planetary companion. We cross-correlated each Keck spectrum against
a synthetic template matching the properties of the star, and averaged
the correlation functions over all orders blueward of the region
affected by the iodine lines. From this representation of the average
spectral line profile we computed the mean bisectors, and as a measure
of the line asymmetry we calculated the ``bisector spans'' as the
velocity difference between points selected near the top and bottom of
the mean bisectors \citep{Torres:05}. If the velocities were the
result of a blend with an eclipsing binary, we would expect the line
bisectors to vary in phase with the photometric period with an
amplitude similar to that of the velocities \citep{Queloz:01,
Mandushev:05}. Instead, we detect no variation in excess of the
measurement uncertainties (see Figure~\ref{fig:rvbis}, bottom panel).
We conclude that the velocity variations are real and that the star is
orbited by a Jovian planet.
	
\section{Star and planet parameters}
\label{sec:parameters}

The mass and radius of the planetary companion scale with those of the
parent star. We have estimated the stellar properties by comparison
with stellar evolution models from the Yonsei-Yale (Y$^2$) series by
\cite{Yi:01}.  However, because a direct estimate of the distance is
unavailable (\hatp3\ was not observed during the {\it Hipparcos}
mission), we have used as a proxy for luminosity ($L_\star$) the
quantity $a/R_\star$ from the light curve fit, which is closely
related to the stellar density \citep[see][]{Sozzetti:07}. As
described by those authors, $a/R_\star$ is typically a better
constraint on $L_\star$ than the spectroscopic value of $\log
g_\star$, which has a relatively subtle effect on the line profiles
and whose determination is therefore more susceptible to systematic
errors. Following \cite{Sozzetti:07} we have determined the range of
stellar masses and radii that are consistent with the measured values
of $T_{\rm eff}$, [Fe/H], and $a/R_\star$ within their
uncertainties. Based on previous experience, for this application we
have adopted more conservative errors for the spectroscopic quantities
of 80~K for $T_{\rm eff}$ and 0.08~dex for [Fe/H].  We obtain $M_\star
= 0.936_{-0.062}^{+0.036}$~M$_{\sun}$ and $R_\star =
0.824_{-0.035}^{+0.043}$~R$_{\sun}$. The predicted surface gravity,
$\log g_\star = 4.577_{-0.045}^{+0.019}$, agrees well with the
spectroscopic estimate (see Table~\ref{tab:stellar}). The nominal age
we infer for \hatp3\ is only 0.4~Gyr, although the uncertainty is very
large due to the unevolved nature of this K dwarf.  While this nominal
age would make \hatp3\ the youngest known transiting extrasolar planet
host star, we note that there is no sign of the
\ion{Li}{1}~$\lambda$6708 line in the spectra of \hatp3\ down to a
factor of $\sim$4 lower than the Li level in stars of similar spectral
type in the Hyades cluster, which have an age of $\sim$0.7 Gyr. This
would suggest an age for \hatp3\ greater than 0.7 Gyr.  Further
evidence for an older age is provided by the measured chromospheric
activity index of the star based on the \ion{Ca}{2} H and K lines in
the Keck spectra: $\logrhk$ is $-4.59\pm0.06$, indicating an age of
$\sim$1.3 Gyr based on the calibration by \cite{noyes84}.

Combined with the results from the light curve and radial velocity
modeling, the above stellar parameters yield a planet mass of $M_p =
0.599 \pm 0.026$~M$_{\rm Jup}$ and a radius of $R_p = 0.890 \pm
0.046$~R$_{\rm Jup}$. The surface gravity of the planet, which is
independent of the stellar parameters \citep[see,
e.g.,][]{Southworth:07}, is $\log g_p = 3.310 \pm 0.066$ (cgs), and is
consistent with the $P$/$\log g_p$ relation shown by those authors.
These and other planet properties are listed in
Table~\ref{tab:parameters}.

The absolute visual magnitude we estimate for the star using the
stellar evolution models ($M_V = 5.86 \pm 0.20$) combined with the
apparent visual magnitude \citep[$V = 11.561 \pm 0.067$;][]{Droege:06}
yields a distance of $140 \pm 13$~pc, ignoring extinction. With this
distance estimate we investigated the possibility that the faint
10\arcsec\ companion to \hatp3\ mentioned in \S\ref{sec:detection} is
physically associated. We find that it is not: its $K_s$-band
brightness as measured by 2MASS is approximately 2 magnitudes fainter
than predicted for a star with its $J\!-\!K_s$ color, using the
best-fit isochrone to \hatp3. It is therefore a background star.

\section{Discussion and concluding remarks}

The properties of \hatp3b\ place it among the smaller transiting
planets discovered so far, and are remarkably similar to those of the
recently announced XO-2b \citep{Burke:07}, not only in mass and radius
(reported as 0.57~M$_{\rm Jup}$ and 0.97~R$_{\rm Jup}$, respectively),
but also in some of the characteristics of the parent star such as
$T_{\rm eff}$ and $M_\star$. In the case of \hatp3b, with a mass of
$M_p = 0.599$~M$_{\rm Jup}$, the radius measured here appears too
small for a pure hydrogen-helium planet according to current theory.
The models for irradiated giant planets by \cite{Fortney:07} indicate
the radius could be explained by the presence of a heavy-element core
of some 75~M$_{\earth}$ ($\sim\frac{1}{3}$ of the total $M_p$) for the
nominal mass, orbital semimajor axis, and age we infer, although
uncertainties in these properties and in the planetary radius allow
the range to be from about 50 to 100~M$_{\earth}$. \hatp3b\ thus joins
HD~149026b \citep{Sato:05} as a hot Jupiter for which theory predicts
a substantial fraction of heavy elements. Using the same models and
the published properties of XO-2b, we find that XO-2b is also expected
to have a 40~M$_{\earth}$ core representing about $\frac{1}{4}$ of its
total mass. These three heavy-element rich planets, along with others
suggested by \cite{Burrows:07}, provide important support for the
core-accretion scenario of planet formation. \hatp3b\ appears
consistent with the correlation proposed by \cite{Guillot:06} between
the host star metallicity and the amount of heavy elements in the
planet interior, although there seems to be at least one exception to
this relation \citep[WASP-1b;][]{Stempels:07}. Compared to HD~149026b,
\hatp3b\ has a similar mass of heavy elements but considerably more
hydrogen and helium, a feature that will need to be explained by
planet formation models.

As of this writing \hatp3b\ is the smallest extrasolar transiting
planet discovered photometrically.  While HD~149026b and GJ~436b
\citep{Gillon:07} do have smaller radii, they were both originally
discovered in Doppler surveys before transits were noticed.

\cite{Gaudi:05} has argued that selection effects in
signal-to-noise-limited transit surveys tend to bias the size
distribution of close-in giant extrasolar planets toward larger radii,
since those objects have deeper transits and are therefore easier to
detect.  The fact that the recent discoveries of \hatp3b\ and XO-2b
are on the small side of the radius distribution would seem to
contradict the argument. However, the transit depth in both of these
cases is not particularly small ($\sim$1.4\%), a result of the fact
that the parent stars are K dwarfs rather than earlier type
stars. Thus, there may still be a bias, although relatively small. We
note also that the properties of \hatp3b\ are entirely consistent with
the Monte Carlo simulations of the population of transiting planets by
\cite{Fressin:07} (see, e.g., their Fig.\ 17), as are other recently
announced transiting planets.  This supports the idea that these
objects are quite representative of the general population of giant
planets, and are not significantly different from the Doppler planets.

\acknowledgements 

We are grateful to the referee, T.\ Guillot, for helpful comments on
the original manuscript.  Operation of the HATNet project is funded in
part by NASA grant NNG04GN74G. We acknowledge partial support also
from the Kepler Mission under NASA Cooperative Agreement NCC2-1390
(D.W.L., PI).  G.T.\ acknowledges partial support from NASA under
grant NNG04LG89G, and work by G.\'A.B.\ was supported by NASA through
Hubble Fellowship Grant HST-HF-01170.01-A.  G.\'A.B.\ also wishes to
acknowledge the use of the photometry software that is under
development by A.\ P\'al. G.K.\ thanks the Hungarian Scientific
Research Foundation (OTKA) for support through grant K-60750. This
research has made use of Keck telescope time granted through NASA, of
the VizieR service \citep{Ochsenbein:00} operated at CDS, Strasbourg,
France, of NASA's Astrophysics Data System Abstract Service, and of
the 2MASS Catalog.

%\clearpage

%\clearpage

%\clearpage

\end{document}